\documentclass[final,5p,times,twocolumn]{elsarticle}

\usepackage{hyperref}
\usepackage{amssymb}
\usepackage{amsmath}
\usepackage{xcolor}
\usepackage{soul}

\usepackage{url}

\journal{Nuclear Inst. and Methods in Physics Research, A}

\begin{document}

\begin{frontmatter}

\title{A Novel Tool for Advanced Analysis of Geant4 Simulations of Charged Particles Interactions in Oriented Crystals}

\author[inst1,inst2]{R. Negrello}

\author[inst2]{L. Bandiera}

\author[inst2]{N. Canale}
\author[inst1,inst2]{P. Fedeli}
\author[inst1,inst2]{V. Guidi}
\author{V. V. Haurylavets}

\author[inst2]{A. Mazzolari}
\author[inst2]{G. Paternò \corref{corr}}
\author[inst2]{M. Romagnoni}
\author{V. V. Tikhomirov}
\author[inst2]{A. Sytov}

\cortext[corr]{Corresponding author; \textit{Email address:} 
\texttt{paterno@fe.infn.it}}

\affiliation[inst1]{organization={Department of Physics and Earth Science, University of Ferrara},
            addressline={Via Saragat 1}, 
            city={Ferrara},
            postcode={44122}, 
            state={Italy}}

\affiliation[inst2]{organization={INFN Section of Ferrara},
            addressline={Via Saragat 1}, 
            city={Ferrara},
            postcode={44122}, 
            state={Italy}}
            

            


\begin{abstract}

We present a novel python tool for the analysis of Geant4 simulations that enhances our understanding of coherent phenomena occurring during the interaction of charged particles with crystal planes. This tool compares the total energy of particles with the potential energy inside crystal channels, enabling a complete examination of coherent effects. By tracking the particle trajectory and classifying the dynamics at each simulation step, it provides deeper insights into how different phenomena contribute to both radiation and particle deflection. This tool can be used to improve crystal-based extraction methods and the development of gamma-ray sources using crystals.
\end{abstract}


\end{frontmatter}

\section{Introduction}
\label{sec:sample1}

Channeling is a well-established phenomenon that occurs when charged particles, such as electrons and positrons, interact with crystalline materials along specific lattice directions. When the particle's incidence angle is less than the critical Lindhard angle\cite{lindhard-1964}, their trajectories become confined within the potential wells formed by the crystal's atomic planes or axes.  For planar channeling, the Lindhard angle is defined as $\vartheta_L = \sqrt{\frac{2U_0}{E}}$, where $U_0$ is the potential well depth and $E$ the particle's energy. Within this angular range, particles experience correlations between successive collisions, enabling a continuous potential approximation for the atomic planes or axes. This confinement induces an oscillating motion, leading to \textit{channeling radiation} \cite{Kumakhov197617} and other coherence effects like \textit{coherent bremsstrahlung} \cite{ferretti1950, ter-mikaelian1972, baryshevskii1976}. Coherent bremsstrahlung occurs when the momentum transferred by the electron or positron to the medium matches a reciprocal lattice vector. These effects are possible both in a straight and in a bent crystal.

Over the past several decades, the phenomenon of channeling in bent crystals has been extensively explored for its potential to steer charged particle beams at several accelerators, including the LHC, Fermilab, and U70 \cite{SCANDALE2016129, PhysRevSTAB.5.043501, Lie, scandale-2013}. Initially proposed by Tsyganov in 1976 \cite{tarantin-1979}, the technique employs planar channeling within bent atomic planes to effectively guide particles along the crystal's curvatures by confining them within the electromagnetic potential wells.

Over the past decades, significant efforts have been dedicated to develop models for the simulation of the dynamics of channeling particles and the emitted radiation. Among the most recent Monte Carlo-based models are RADCHARM++\cite{guidi-2012, bandiera-2015}, CRYSTALRAD++\cite{sytov-2019}, MBN Explorer\cite{sushko-2013, solovyov2012}, and the Geant4 G4ChannelingFastSim Model \footnote{Further details about the Channeling Fast Sim Model can be found at \url{https://geant4-userdoc.web.cern.ch/UsersGuides/PhysicsReferenceManual/html/solidstate/channeling/channeling_fastsim.html}.} \cite{sytov-2023}, which implements the physics of channeling directly within the Geant4 toolkit \cite{Agostinelli2003, Allison2006, Allison2016}.  This last model provides a comprehensive environment for simulating channeling processes, and the classification tool presented in this work was specifically designed to analyze and classify the simulation outputs it generates.

In addition to simulation models, preliminary methods for classifying particle dynamics in crystals have been developed. Earlier works \cite{sytov-2017, mazzolari-2014, PhysRevLett.115.025504, korol2014, korol2017, korol2021} introduced foundational frameworks for understanding the relation between particle behavior within crystalline potentials and radiation emitted. These approaches provided valuable insights and were furtherly developed in this study.

This work presents a detailed Python-based framework designed to analyze specific outputs from Geant4 simulations, focusing on particle deflection and radiation processes in bent crystals. 

\section{Methodology for Particle and Photon Classification}
The Geant4 simulation output files are analyzed step by step for each particle trajectory. At each step, the particle's state is determined based on its interaction with the crystal potential, enabling a detailed classification of its dynamics throughout the trajectory.

\subsection*{State Determination}
The classification of particle states (\textit{channeling} or \textit{over barrier}) relies on comparing the transverse energy of the particle with the effective potential of the crystal. This process involves the following steps:
\begin{figure*}
    \centering
    \includegraphics[width=0.8\linewidth]{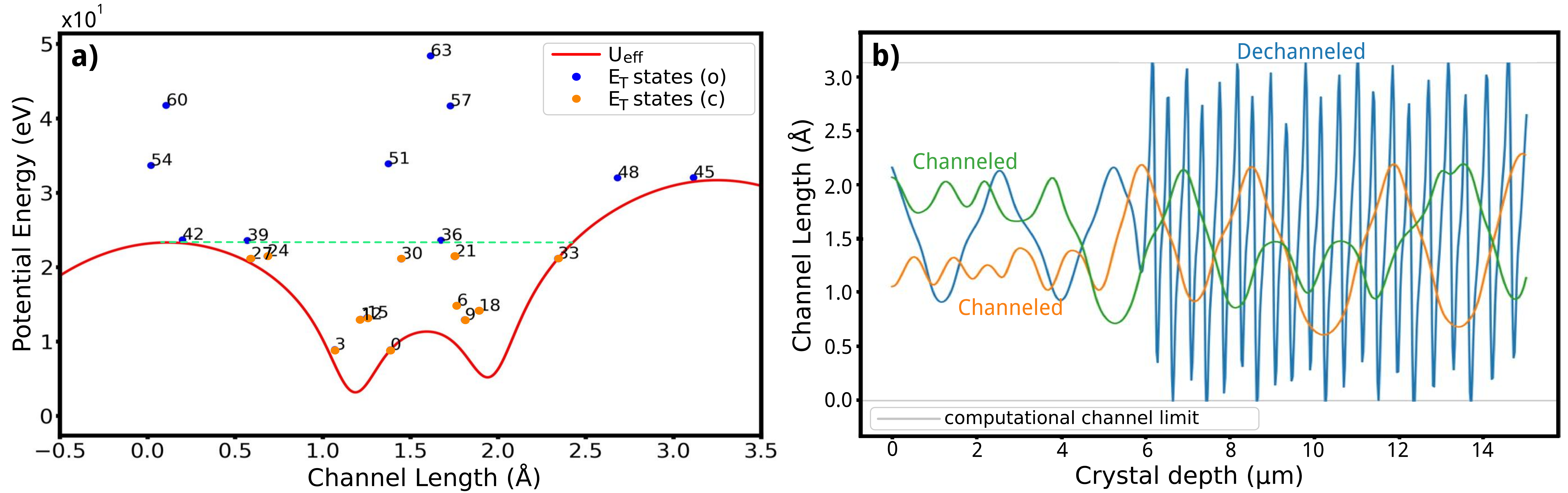}
    \caption{(a) Potential energy versus channel length for a bent crystal. The red curve represents the effective potential $U_{\text{eff}}$, while the green dashed line indicates the channel boundary. The particle states are shown as discrete points: orange dots correspond to channeling states, and blue dots correspond to overbarrier states. For clarity, only one out of every three saved states is plotted.  (b) Particle trajectory corresponding to channeling and dechanneling case. The trajectory is harmonic in the channeling state but crosses several channels upon dechanneling.}
    \label{fig:potandtraj}
\end{figure*}
\begin{enumerate}
    \item \textbf{Input Data Extraction:} \\
    For each particle and simulation step, the position ($x$) and the direction ($\vartheta_x$) relative to the channel are extracted from Geant4 simulation output files. In the Geant4 simulation code, the trajectory of the particle is obtained by solving the equation of motion for a particle in a potential field using Runge-Kutta method. Samples of particle trajectory and direction are saved after a variable number of steps that can be choosen after some iterations, typically for short crystal the step is of the order of tens of $nm$. These quantities are essential for calculating the transverse energy.
    \item \textbf{Effective Potential Calculation:}\\
    he effective potential for the bent crystal is calculated at each particle position by combining the interatomic potential and the bending contribution, as described by the Doyle-Terner model\cite{DUDAREV199586}:
    \begin{equation}
    U_{\text{eff}}(x) = U_{\text{straight}}(x) + \frac{pv}{R} x,
    \end{equation}
    where $U_{\text{straight}}(x)$ represents the interatomic potential for a straight crystal, $p$ and $v$ are the particle's momentum and velocity, respectively, and $R$ is the bending radius of the crystal.

    \item \textbf{Calculation of the Transverse Energy ($E_T$):}\\
    The transverse energy at each trajectory point $(x, \vartheta_x)$ is computed as \cite{Biryukov1997}:
    \begin{equation}
        E_T = \dfrac{pv}{2}\vartheta_x^2 + U_{\text{eff}}(x),
    \end{equation}
    where $\vartheta_x=\dfrac{dx}{dz}$ is the incident angle with respect to the atomic plane, and $U_{\text{eff}}(x)$ is the effective potential at position $x$ previously calculated.

    \item \textbf{Comparison with the Effective Potential:}\\
    For each particle step, the calculated transverse energy $E_T$ is compared with the maximum value of the effective potential $U_{\text{eff}}(x)$. A state is assigned as follows: \textbf{Channeling (c):} $E_T \leq U_{\text{eff,max}}$ or \textbf{Overbarrier (o):} $E_T > U_{\text{eff,max}}$.
\end{enumerate}
Fig.~\ref{fig:potandtraj} a) illustrates the classification process. The red curve represents the effective potential $U_{\text{eff}}$, with the green dashed line marking the channel boundary. Particles with transverse energy $E_T$ below this limit are classified as channeling (\textit{c}), while those above are overbarrier (\textit{o}). Orange and blue points represent channeling and overbarrier states, respectively, along the trajectory. The transition from channeling to overbarrier demonstrates the dechanneling process. 

Given the large number of particle trajectories and steps, the calculation of $E_T$ and state determination is parallelized.
\subsection*{Trajectory Classification}

A dedicated algorithm has been developed to classify particles based on their interaction dynamics within the crystalline potential. This function evaluates the sequence of states along each particle's trajectory, identifying characteristic patterns that correspond to distinct physical phenomena. The classification is performed by analyzing the state transitions throughout the trajectory.

The classification includes the following cases:

\begin{enumerate}
    \item \textbf{Channeling:} 
    The particle remains confined within the potential channel throughout its entire trajectory. This occurs if all states are channeling (\textit{c}).

    \item \textbf{Overbarrier:} 
    The particle moves entirely above the potential barriers without entering a channeling state. This occurs if all states are overbarrier (\textit{o}).

    \item \textbf{Dechanneling:} 
    The particle starts in channeling and transitions to overbarrier, remaining in that state thereafter.

    \item \textbf{Captured Overbarrier:} 
    The particle begins in overbarrier, transitions into channeling, and remains confined for the rest of its trajectory.

    \item \textbf{Captured Overbarrier with Dechanneling:} 
    The particle starts in overbarrier, transitions into channeling, but subsequently undergoes dechanneling and remains in overbarrier thereafter.

    \item \textbf{Rechanneling:} 
    The particle undergoes multiple transitions between channeling and overbarrier states during its trajectory. The number of transitions is counted and included in the classification. For example:\textit{rechanneling} indicates one transition to channeling followed by remaining confined. \textit{rechanneling and dechanneling} indicates one transition to channeling followed by dechanneling. \textit{Multiple rechanneling} events are considered by counting the number of transitions.

\end{enumerate}
Fig.~\ref{fig:potandtraj} b) depicts two channeled trajectories and one dechanneled emerging from the classification. To simplify calculations, the code assumes periodic boundary conditions within a single channel (gray lines), causing dechanneled particles exiting one boundary to reappear at the opposite, resulting in the observed trajectory behavior after 6 $\mu$m.

\subsection*{Photon Classification}

The radiation emission model implemented in the G4ChannelingFastSim Model relies on the direct integration of the general Baier-Katkov formula for calculating electromagnetic radiation emitted by charged particles in an external field \cite{PhysRevA.86.042903}. This approach enables precise computation of the probability of radiation emission during particle trajectories. As mentioned in the previous subsection, the particle trajectory is calculated by integrating step by step the equation of motion. At each step, the probability of emitting a photon is accumulated by integrating the Baier-Katkov formula up to the actual trajectory point to build the Cumulative Distribution Function (CDF).
After a given number of trajectory steps (by default 10000) or if at some point the photon emission probability exceeds a threshold value (by default 0.25), we simulate whether radiation actually takes place by drawing a random number $r$ uniformly distributed in the interval [0,1]. If $r$ is smaller than the calculated probability, a photon is generated at the point where the CDF equals $r$; otherwise, no photon is emitted. The CDF is then reset, and the trajectory simulation continues from either the emission point or the last trajectory point if no emission occurred.

Building upon this foundation, the complementary classification process developed in this work incorporates emitted photons into the particle trajectory analysis. This extension allows for a detailed examination of radiation contributions by distinguishing between particles that emit photons and those that do not. The methodology consists of the following steps:

\begin{enumerate}
    \item \textbf{Separation of Particles by Emission Behavior}\\
    Particles are divided into two groups: \textit{Radiating Particles:} Particles that emit one or more photons during their trajectories. These are identified by analyzing the photon spectrum extracted from the simulation files. \textit{Non-Radiating Particles:} Particles that traverse the crystal without emitting photons. These are processed separately to maintain a comprehensive classification.

    \item \textbf{Sub-Trajectory Extraction}\\
    When a photon is emitted, the particle trajectory is segmented into sub-trajectories. The trajectory up to the emission point is treated as one segment,  each subsequent photon emission results in additional sub-trajectories. This segmentation ensures that the photon is correctly associated with the phenomena the particle was experiencing when it was emitted.

    \item \textbf{Photon State Classification}\\
    For each sub-trajectory, the states are re-evaluated using the same potential and transverse energy calculations applied to the full trajectories.
    A secondary classification is performed to determine the behavior of the particle during the emission process. In this way we have both the classification of the particle though the full trajectory and the one until the emission point.

\end{enumerate}

\begin{figure} \centering \includegraphics[width=0.8\linewidth]{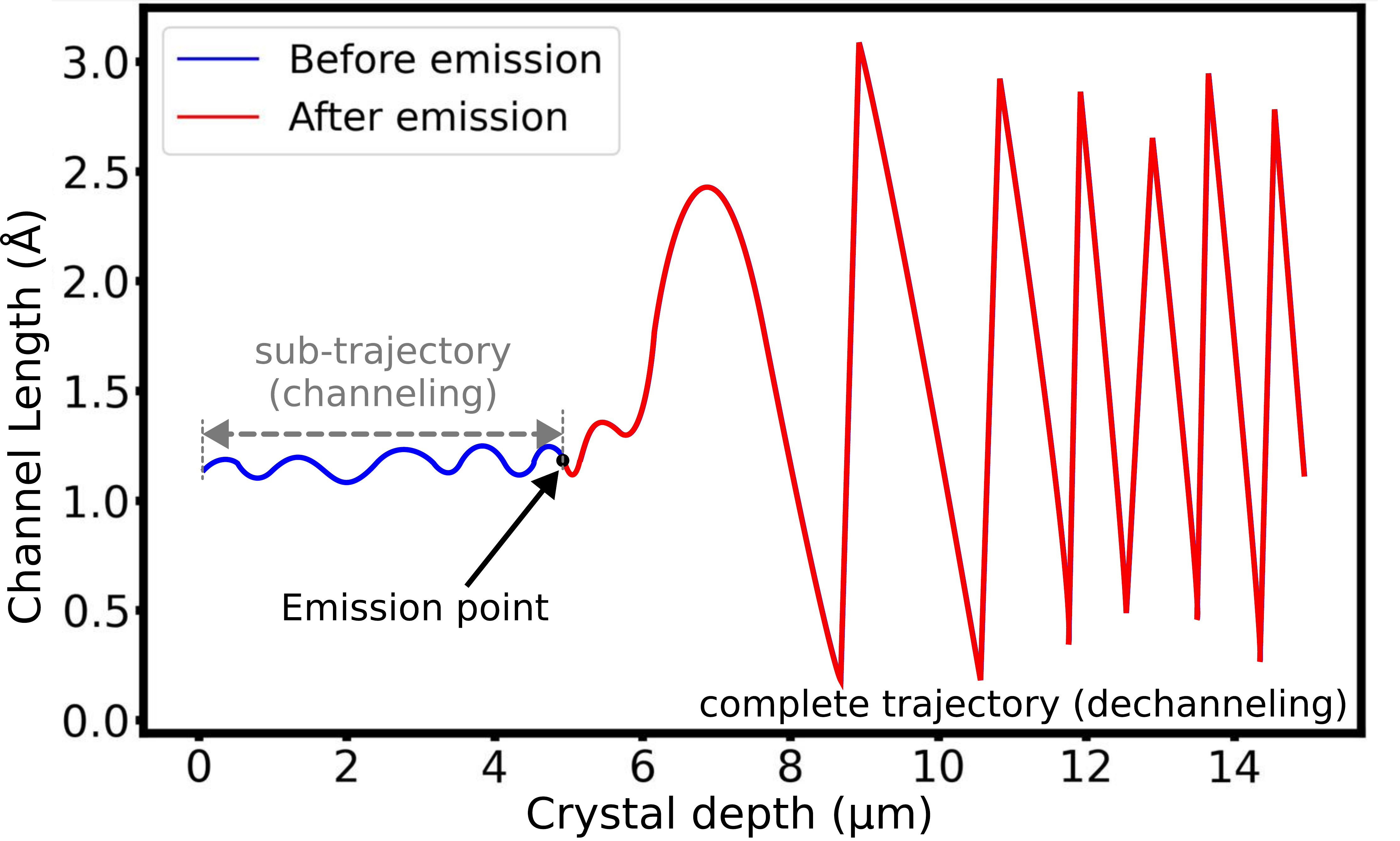} \caption{Visualization of the trajectory segmentation process during photon emission. The particle starts in a channeling state (blue curve) and transitions to a dechanneling state after emission (red curve). The emission point is marked with a black arrow. The sub-trajectory before emission is analyzed separately to classify the phenomena responsible for the emission.} \label{fig:emission_trajectory} \end{figure}

The particle trajectory segmentation process is visually illustrated in Fig.~\ref{fig:emission_trajectory}. Before emission, the particle is confined to the channeling regime, following an oscillatory trajectory within the crystal potential well. At the emission point, the particle emits a photon and undergoes a momentum kick, transitioning into a dechanneling state transversing some channels. This segmentation is critical for isolating the contribution of specific phenomena during the emission process.
The information from particle trajectory classification, photon emissions, and sub-trajectories is merged to form a comprehensive dataset.

\section{Simulation results and discussion}
In this section the main output of the analysis process are shown and discussed, initially considering the deflection pattern and then focusing on the radiation emitted.

\subsection*{Analyzing the Deflection Pattern}

\begin{figure}
    \centering
    \includegraphics[width=1\linewidth]{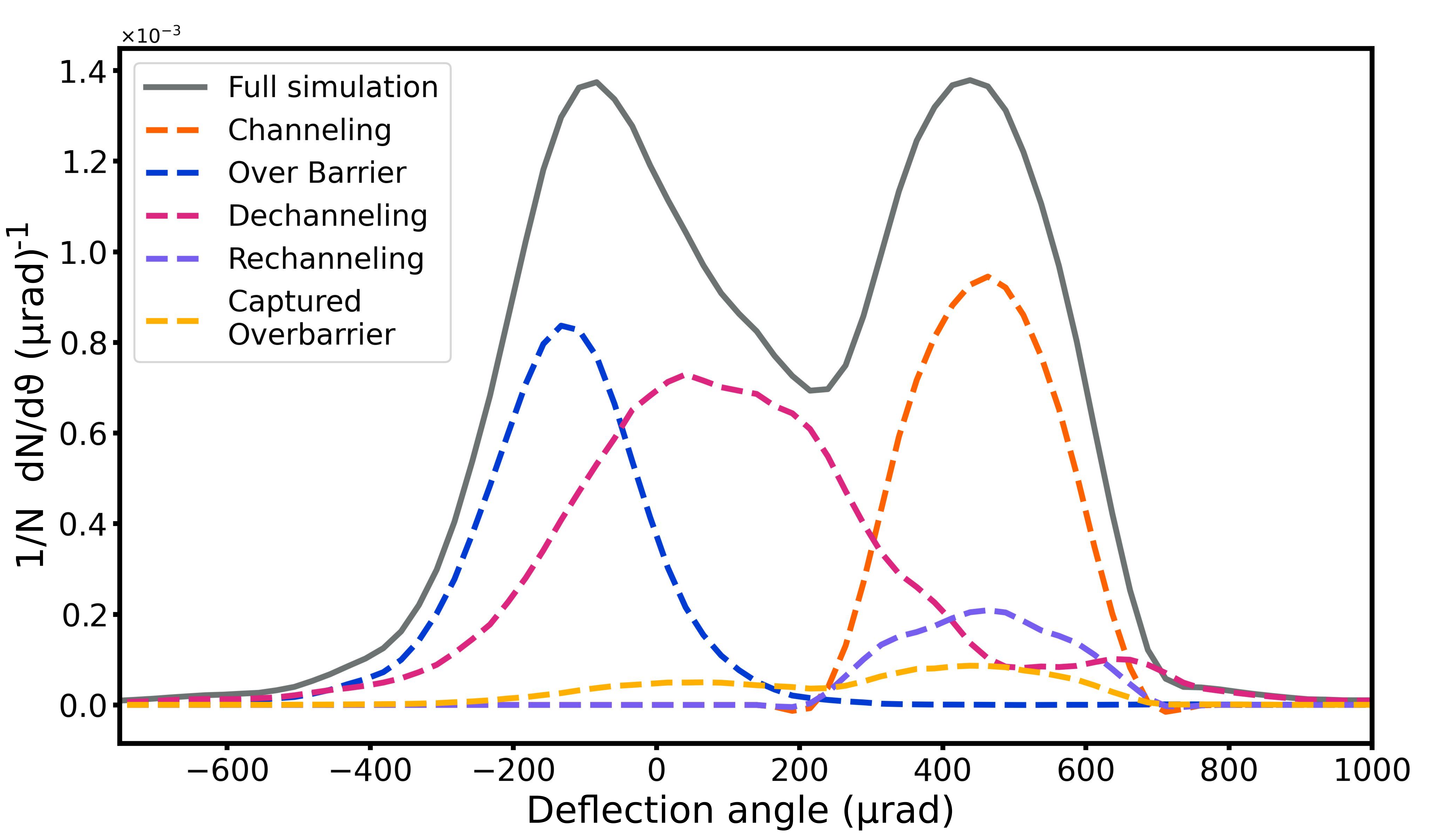}
    \caption{Deflection pattern of a simulated electron beam interacting with a silicon crystal. The total simulated deflection is shown as a solid dark gray line, with contributions from individual processes identified using the classification algorithm: channeled particles (orange dashed); over barrier particles (blue dashed); dechanneled particles (magenta dashed); rechanneling particles (purple dashed); captured over barrier particles (ochra dashed).}
    \label{fig:classification_output}
\end{figure}

The particle classification algorithm enables a detailed analysis of the deflection patterns produced by particle beams interacting with crystals, providing insight into the contributions of various physical processes. Figure \ref{fig:classification_output} illustrates the simulated deflection pattern for a $15\,\ \mu \text{m}$-thick silicon crystal aligned with the (111) atomic planes and bent to an angle of $455\,\  \mu \text{rad}$. An electron beam of $855\,\text{MeV}$ impinges on the crystal at an incident angle centered around zero. The beam follows a Gaussian spatial distribution with dimensions of $50  \, \mu \text{m} \times 100 \, \mu \text{m}$ in $(x,y)$ and a divergence of $30 \,\mu \text{rad}$ along both axes.
The simulation shows two distinct peaks: the left peak represents the half-reflected beam, and the right peak corresponds to the deflected beam (solid dark gray line). The classification framework identifies the contributions of specific processes, represented by dashed lines: channeling (orange), overbarrier (blue), dechanneling (magenta, including rechanneled particles that dechannel again), rechanneling (purple, categorized by the number of rechanneling events), and captured overbarrier (ochre, transitioning into either channeling or dechanneling).

This detailed analysis is crucial for understanding the underlying dynamics of channeling and deflection processes in crystals. It provides valuable insights into optimizing these processes for specific applications, such as beam steering or enhancing radiation production in particle accelerators.

\subsection*{Spectral Analysis of Emitted Radiation}

\begin{figure}[!h]
    \centering
    \includegraphics[width=1\linewidth]{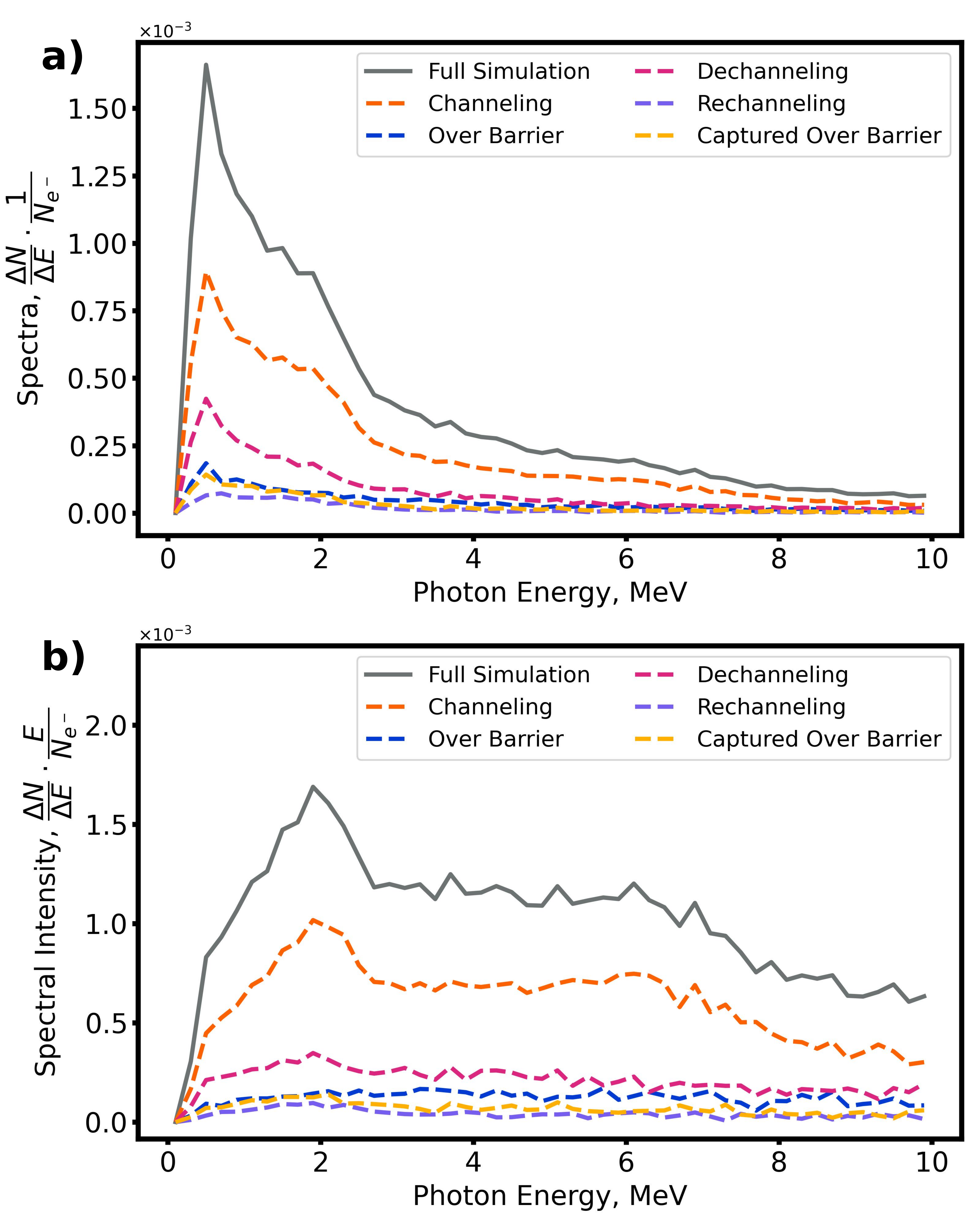}
    \caption{a) shows the photon energy spectrum normalized to the total number of particles and bin size, while b) displays the spectral intensity normalized in the same manner. The full simulation is shown as a dark gray solid line. The contributions from different particle states are distinguished: channeling (orange dashed), dechanneling (magenta dashed), overbarrier (blue dashed), captured overbarrier (ochre dashed), and rechanneling (purple dashed).}
    \label{fig:spectrum_analysis}
\end{figure}

Figure~\ref{fig:spectrum_analysis} highlights the classification framework's capability to disentangle the spectral contributions of different particle states to the emitted photon spectrum and spectral intensity. Simulations were conducted using the same crystal and beam of the previous section, considering only photon energies above 200 keV. Figure~\ref{fig:spectrum_analysis}a presents the photon energy spectrum, while Figure~\ref{fig:spectrum_analysis}b) depicts the spectral intensity. The dark gray solid line shows the total spectrum, with individual contributions identified: channeling (orange dashed) dominates, followed by dechanneling (magenta dashed), overbarrier (blue dashed), captured overbarrier (ochre dashed), and rechanneling (purple dashed). This classification provides valuable insights into the radiation emission mechanisms in bent crystals.

This analysis confirms that channeling contributes approximately 60\% to the total spectrum, making it the most significant phenomenon. Additionally, the code enables the quantification of emission efficiency: about 1.1\% of all particles in the channeling state emit photons. Such detailed classification can be used for investigating the underlying physics of channeling and optimizing crystal configurations for specific applications.

\section{Conclusions}

This work presents a novel classification framework designed to enhance the analysis of Geant4 simulations, providing an improved understanding of the coherent effects in bent crystals. By integrating particle state classification with photon emission dynamics, the framework allows the study of both deflection patterns and radiation spectra. This capability has been demonstrated through detailed analysis, revealing distinct contributions from the different phenomena. For example, it was possible to quantify that approximately 60\% of the radiation spectrum is attributed to channeling, while only 1.1\% of channeling particles emit photons.
The significance of this advancement lies in its ability to quantify contributions from various particle states with improved precision. This tool has the potential to help the design and optimization of crystal-based systems, including applications in particle accelerators, beam steering, and gamma-ray sources.

\section*{Funding}
\small

\noindent{This work was supported by INFN CSN5 (OREO and GEANT4INFN projects) and the European Commission through the H2020-MSCA-RISE N-LIGHT (G.A. 872196), EIC-PATHFINDER-OPEN TECHNO-CLS (G.A. 101046458) and  MHz-TOMOSCOPY (G.A 101046448) projects. A. Sytov acknowledges support from the H2020-MSCA-IF-Global TRILLION (G.A. 101032975) and the European Union – NextGenerationEU – Project Title : "Intense positron source Based On Oriented crySTals - e+BOOST" 2022Y87K7X – CUP I53D23001510006.}

\bibliography{biblio}

\end{document}